\documentclass[global,twocolumn]{svjour}
\usepackage{graphicx}
\journalname{Appl.~Phys.~B}

\begin{document}

\newcommand{\dip}[1]{\cdot 10^{#1}}

\newcommand{\vecx}{\mathbf{e}_x}
\newcommand{\vecy}{\mathbf{e}_y}
\newcommand{\vecz}{\mathbf{e}_z}

\newcommand{\abs}[1]{\left|#1\right|}

\newcommand{\vect}[1]{\mbox{\bf #1}}
\newcommand{\sub}[1]{_{\mbox{\scriptsize #1}}}
\newcommand{\hyp}[1]{^{\mbox{\scriptsize #1}}}
\newcommand{\bigabs}[1]{\bigl|#1\bigr|}
\newcommand{\mikrom}{\mbox{$\mu$m}}
\newcommand{\ket}[1]{\mbox{$\left|#1\right\rangle$}}

\newcommand{\I}[1]{I_{\textrm{#1}}}

\title{Applications of Integrated Magnetic Microtraps}
\author{J.~Reichel\thanks{\emph{Fax:} +49\,89\,285192; \emph{e-mail:} jakob.reichel@physik.uni-muenchen.de}
 \and W.~H\"ansel \and P.~Hommelhoff
  \and T.~W.~H\"ansch
}                     
\institute{Max-Planck-Institut f\"ur Quantenoptik and Sektion Physik der
  Ludwig-Maximilians-Universit\"at\\Schellingstr. 4, D-80799
  M\"unchen, Germany}
\date{Submitted 24.07.2000, revised version 25.08.2000}
%

\maketitle
\begin{abstract}
  Lithographically fabricated circuit patterns can provide magnetic
  guides and microtraps for cold neutral atoms. By combining several
  such structures on the same ceramic substrate, we have realized the
  first ``atom chips'' that permit complex manipulations of ultracold
  trapped atoms or de Broglie wavepackets. We show how to design
  magnetic potentials from simple conductor patterns and we describe
  an efficient trap loading procedure in detail.  Applying the design
  guide, we describe some new microtrap potentials, including a trap
  which reaches the Lamb-Dicke regime for rubidium atoms in all three
  dimensions, and a rotatable Ioffe-Pritchard trap, which we also
  demonstrate experimentally. Finally, we demonstrate a device
  allowing independent linear positioning of two atomic clouds which
  are very tightly confined laterally. This device is well suited for
  the study of one-dimensional collisions.
\end{abstract}


\section{Introduction}
\label{intro}
Trapped laser cooled atoms are now widely used in atomic physics
experiments \cite{Ketterle99a}. Magnetic traps are particularly
versatile, since they can be used for any atomic species possessing a
magnetic moment, and provide conservative potentials even for long
trapping times.  Although the magnetic potential has merely served as
a container in most experiments to date, it also provides an excellent
means for more complex manipulation of atomic wave packets. In
resonator QED experiments \cite{Pinkse00,Ye99} for example, it can be
used to shift a trapped atom into and out of the quantized light
field. Quantum computing schemes have been proposed in which a
combination of electric and magnetic potentials controls
state-dependent collisions in order to entangle atomic qubits
\cite{Calarco00}.  Such complex potentials can be realized when
lithographic or other surface-patterning processes are used to produce
the field-creating structures on a suitable substrate, now sometimes
called an ``atom chip''.  The use of lithographic, planar conductor
patterns for magnetic atom trapping has been proposed as early as 1995
\cite{Weinstein95}. However, no successful experiments were carried
out at that time due to the difficulty of loading such traps, which
have small volumes and are typically located only a few hundred
micrometers or less from the substrate surface.  This situation
changed last year with our demonstration of an efficient loading
mechanism for surface traps \cite{Reichel99}. It employs a novel
mirror-MOT, using a reflecting layer on top of the circuit pattern to
realize the laser fields for laser cooling and trapping in close
proximity to the surface. With this loading mechanism, integrated
traps for cold atoms became experimentally accessible.  In the first
demonstration of such a trap, the new loading mechanism was employed
to fill a miniature quadrupole trap featuring transverse gradients of
1700\,G$/$cm \cite{Reichel99}.  The trapping potential was created by
a lithographically produced, U-shaped conductor in conjunction with an
external bias field. In the meantime, the same loading and trapping
techniques have also been applied to construct a Ioffe-Pritchard (IP)
trap for $^7$Li atoms \cite{Folman00}. The mirror-MOT technique has
also proven its usefulness beyond lithographic traps as it was used to
fill surface traps of a different type, which result from the
combination of permanent surface fields on magnetic tape with an
external bias field \cite{Rosenbusch00}.

Indeed, lithographic conductors are now increasingly employed to trap,
guide and manipulate cold neutral atoms. Parallel conductors have been
used to realize atom guides in which atoms are confined in two
dimensions and move freely along the third \cite{Mueller99,Dekker00}.
Very recently, more complex conductor patterns were used to realize an
atomic conveyer belt, which adiabatically transports atoms and
positions them with a precision of the order of $1\,\mu$m
\cite{Haensel00a}. Moreover, two ``beam splitters'' have been
demonstrated, one which distributes a trapped cloud in a Y-shaped
pattern \cite{Cassettari00}, and one which splits a guided beam using
an X-shaped pattern \cite{Mueller00}. Such devices demonstrate the
versatility of the new approach. Considering the relative ease of
these experiments now that major obstacles have been removed, we
foresee a wide variety of applications for lithographic microtraps. In
this article, we discuss some key issues of microtrap design and
realization. We propose a modular, intuitive approach to the design of
complex potentials from a few simple building blocks.  In the
following sections, we tackle the substrate technology and give a
detailed description of the loading procedure. The wide-ranging
possibilities offered by simple conductor configurations are
illustrated by experimental results of trapping in some fundamental
types of potentials which would be very difficult to create by more
traditional means.  Finally, we demonstrate a linear collider for
trapped atoms, which is well suited for the study of one-dimensional
collisions.

\section{Design of integrated microtraps}
\label{sec:design}

In this section we present a method to design a variety of microtraps
from simple 2D-conductor configurations which serve as modular
building blocks. As central part we investigate the magnetic potential
at a perpendicular wire intersection, which can then be used to
construct more complex magnetic potentials. Throughout the discussion
we will denote $\vect{e}_z$ the vector normal to the substrate
surface, all current carrying wires are contained within the $xy$
plane at $z$=0.

\subsection{2D-quadrupole fields}
\begin{figure}[h]
  \includegraphics[width=0.95\columnwidth]{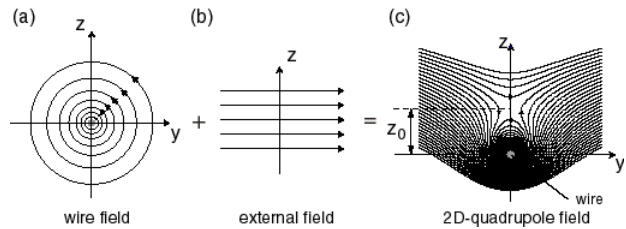}
\caption{The field of a current carrying wire (a) and a homogenous
external field (b) add up to a 2D-quadrupole field above the wire
(c) .} \label{fig:WireQuadrupole}
\end{figure}

Ioffe-Pritchard (IP) traps can be regarded as superposition of a
two-dimensional quadrupole field for transverse confinement and a
longitudinally varying field for confinement along the quadrupole
axis. The 2D-quadrupole field can easily be created in the vicinity of
a current carrying wire (here $I_0$ along $\vect{e}_x)$, if its
tangential field $B=\frac{\mu_0}{2\pi}\,\frac{I_0}{R}$ is compensated
at the point $\vect r=(x,0,z_0)$ by the homogeneous field
 \begin{equation}
 \vect B = B_{0,y}\,\vect{e}_y = \frac{2\,\mbox{G}}{z_0 /
 \mbox{mm}}\,I/\mbox{A}
 \end{equation}
(see fig.\,\ref{fig:WireQuadrupole}).

Magnetic traps and waveguides employing this scheme have been realized
both with discrete \cite{Fortagh98,Denschlag99} and with lithographic
wires \cite{Reichel99,Dekker00,Folman00}.  Furthermore, schemes have
been studied that permit the creation of quadrupole guides without the
use of external fields (see e.g.  \cite{Thywissen99}).

\subsection{Modification of the longitudinal potential}
The field strength along the longitudinal trap axis can be varied
by adding another field $\tilde{\vect{B}}(\vect{r})$ to the
quadrupole field $\vect B\hyp Q$. The components
$\tilde{B}_y,\tilde{B}_z$ yield a shift of the
trap position along $\vect{e}_z$ and $\vect{e}_y$ respectively,
whereas the modulus of $\tilde{B}_x$ determines the field strength
at the shifted minimum location. Using a linear approximation for
the quadrupole field
 \begin{eqnarray}
 \vect{B}\hyp{Q}(x,y,z)&=&(0,-b\,(z-z_0),b\,y) \\
 b&=&\frac{2\pi}{\mu_0}\,\frac{B_{0,y}^2}{I_0}\,,
 \end{eqnarray}
the total field $\vect{B}={\vect{B}}\hyp{Q} + \tilde{\vect{B}}$
and the transverse trap position can be expressed to
0$^{\mbox{th}}$ and 1$^{\mbox{st}}$ order, respectively, as
\begin{eqnarray}
 B\sub{min}(x)&=&
 \abs{{\tilde
 B}_x(x,y=0,z_0)}+ O(\frac{\tilde B_y}b,\frac{\tilde B_z}b)
 \label{eq:BMin}\\
 {\vect r}\sub{min}(x) &=&
 (x,-\frac{{\tilde{B}}_z}b,z_0+\frac{{\tilde{B}}_y}b) +
 O(\bigl(\frac{\tilde B_y}b\bigl)^2,\bigl(\frac{\tilde B_z}b\bigl)^2).\label{eq:rMin}
\end{eqnarray}
Here $B\sub{min}(x)$ signifies the lowest value of the magnetic field
within the plane perpendicular to the $x$~axis, $r\sub{min}$ denotes
the corresponding minimum position. According to eq.\,\ref{eq:BMin}
the transverse field minimum can be approximated by $\abs{\tilde
  B_x(x,0,z_0)}$ taken at the position of the unshifted quadrupole
guide. This method can be used to develop an intuitive understanding
of many magnetic potentials that are based on a 2D-quadrupole trap.

\subsection{Wire intersection}
\label{sec:intersection}

\begin{figure}
    \centering
 \includegraphics[width=\columnwidth]{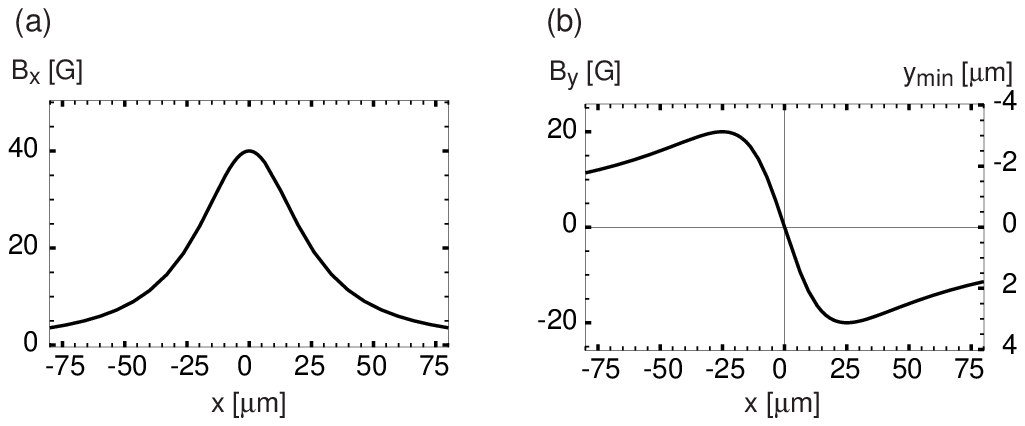}
 \caption{Field contributions of an intersecting wire at height $z_0$=25\,$\mu$m
 for a current of $I_1$=0.5\,A.}
 \label{fig:BXY}
\end{figure}

A high field ${\tilde B}_x$,
i.e.\ a strong longitudinal potential, may be obtained by a wire
that intersects the x axis at right angle.
\begin{figure}[h]
\centering
  \includegraphics[height=4cm,width=\columnwidth,keepaspectratio]{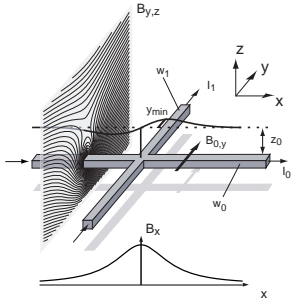}
 \caption{The field strength in the center of the 2D-quadrupole can be modified
   by a current crossing the first wire perpendicularly. The x
   component of its magnetic field determines the longitudinal trap
   potential (bottom), the y component displaces the transverse field
   minimum (solid line) from the the original quadrupol line (dashed)
   in the xy plane.}
 \label{fig:Intersection}
\end{figure}
Fig.\,\ref{fig:Intersection} illustrates the geometry of the wire
intersection that will serve as building block for more
complicated traps. For demonstration, we choose a current of
$I_0=2$\,A which together with a magnetic bias field of
$B_{0,y}=160$\,G yields a 2D-quadrupole guide $z_0=25\,\mikrom$
above the surface\footnote{These parameters are realistic, see
Sect.~\ref{sec:substrate}}. At this distance, the current $I_1$ creates a
field $\vect{B}\hyp{w}$ with components
 \begin{eqnarray}
 B\hyp{w}_x&=&\frac{\mu_0}{2\pi}\,\frac{I\,z_0}{x^2+z_0^2} \label{eq:BWX} \\
 B\hyp{w}_y&=& 0 \label{eq:BWY}\\
 B\hyp{w}_z&=&-\frac{\mu_0}{2\pi}\,\frac{I\,x}{x^2+z_0^2} \label{eq:BWZ}
 \end{eqnarray}
 which are shown in
fig.\,\ref{fig:BXY}.

\begin{figure}
    \centering
  \includegraphics[width=\columnwidth,height=7cm,keepaspectratio]{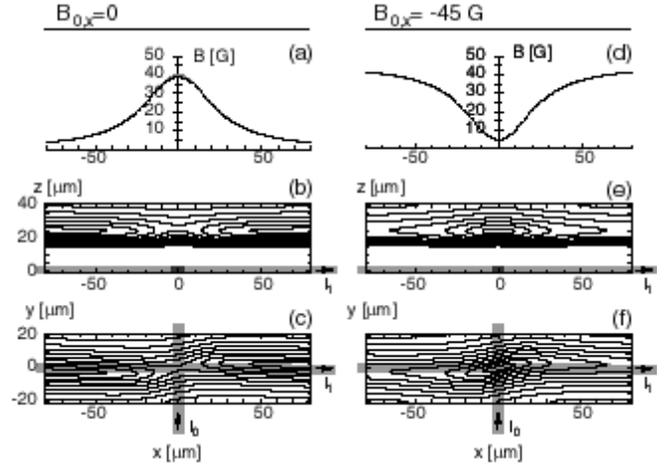}
 \caption{Field strength at a single wire intersection, parameters are $I_0$=2\,A,
 $I_1$=0.5\,A, $B_{0,y}$=160\,G, contour lines mark the field strength in multiples of 10\,G.
 Without an additional
 field along $\vect{e}_x$ the crossing current leads to a repelling potential (left column),
 whereas it creates a three-dimensional trap
 above the intersection point if an additional field
 $B_{0,x}$=\mbox{-45\,G}$~\vect{e}_x$ is applied (right column).}
 \label{fig:ContoursC}
\end{figure}

As the lorentzian shape of $\vect B\hyp{w}_x$ suggests, the field
contribution of the intersecting wire can be used to create a
repulsive potential along the quadrupol axis
(Fig.\,\ref{fig:ContoursC} a-c). However, the repulsive character
of the crossing current can be converted into an attractive one if
a homogeneous bias field $B_{0,x}< - B\hyp{w}_x(x=0,z_0)$ is
superimposed. In this case, the field along the axis
becomes
\begin{equation}
 B\sub{min}(x)\approx\bigabs{\tilde{B}_x(x,y=0,z_0)}=\abs{B_{0,x}} -
 \abs{B\hyp{w}_x(x,z_0)}\,
\end{equation}
which exhibits a minimum at $x=0$. This field configuration thus
provides a trapping potential in all three dimensions
(fig.\,\ref{fig:ContoursC} d-f). Wires supporting a current flow
perpendicular to a quadrupole guide can therefore be used to
create repulsive as well as attractive potentials along the trap
axis, depending on the strength of the external field $B_{0,x}$.

In order to reflect the physical properties of the trapping
potentials, each plot in fig.\,\ref{fig:ContoursC} shows the
minimum potential value within the plane (a,d) or line (b,c,e,f)
that is perpendicular to the visible line or plane.

A detailed look at the potentials in fig.\,\ref{fig:ContoursC}
reveals how the minimum position of the trap is displaced by the
field component $B\hyp w_y$ of the crossing current. The
approximations in eq.\,\ref{eq:BMin} and \ref{eq:rMin} hold as
long as the intersecting current is smaller than the current in
the central wire ($I_1\ll I_0$). Here we have chosen $I_1=\frac14
I_0$, the approximation therefore predicts $y\sub{min}$ to vary
according to
\begin{equation}
 y\sub{min}=-b\,B\hyp w_z = -6.4\,\mikrom\,B\hyp w_z / \mbox{G}\,,
\end{equation}
as is indicated on the right hand scale of
fig.\,\ref{fig:BXY}~(b). The deviation from the approximated field
strength is indeed so small that it is not visible on the scale of
the plots (a) and (d) of fig.\,\ref{fig:ContoursC}.

\subsection{Composing more complex traps}
\begin{figure}
    \centering
  \includegraphics[width=\columnwidth,height=7cm,keepaspectratio]{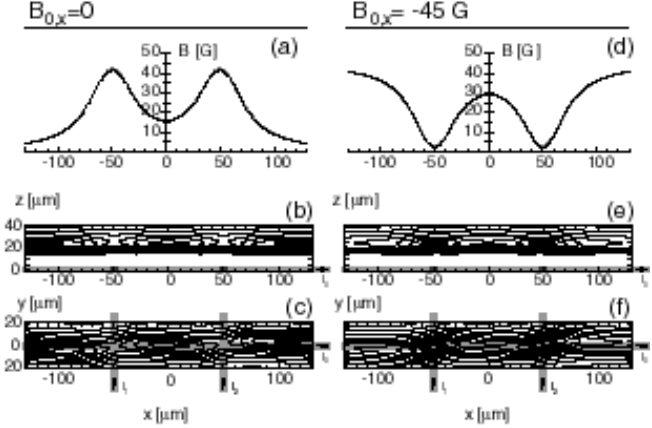}
 \caption{Field strength at two neighbouring wire intersections. The 
   distance of the crossing wires has been chosen $d=4z_0=100\,\mu$m,
   parameters are $I_0$=2\,A, $I_1$=$I_2$=0.5\,A, $B_{0,y}$=160\,G.
   $B_{0,x}$ is chosen as in fig.\,\ref{fig:ContoursC}.}
 \label{fig:ContoursH}
\end{figure}

The fields of several intersecting wires can now be added to
obtain more complex magnetic potentials. Fig.\,\ref{fig:ContoursH}
illustrates the potentials that arise in the vicinity of two
parallel wires that intersect the central wire at two points
100\,\mikrom\ apart from each other. For the intuitive
understanding the field contributions from the two wire can be
summed up in order to give the complete longitudinal potential.

Without external bias field $B_{0,x}$, the field configuration
exhibits one IP-trap in the center of the two intersecting wires.
Contrary to conventional IP-traps, the field direction in the
center does not coincide with the slowest trapping axis: the
center field is along $\vect{e}_x$, whereas the slow axis is
turned by the influence of the vertical field components of the
intersecting wires.

If a bias field $B_{0,x}$ is applied, two IP-traps form above the
intersection points, each slightly displaced by the influence of the
opposite wire (see fig.\,\ref{fig:ContoursH} f). In a similar way, a
long chain of magnetic traps can be constructed. The magnetic conveyer
belt \cite{Reichel99,Haensel00a} can be understood in terms of this
design principle. As multiple intersections of the modulating wires
cannot be realized in a single-layer substrate, the wires are bent off
to avoid the crossing.

\subsection{Choosing configuration parameters}
The analytical form of the longitudinal trap potential
(eq.\,\ref{eq:BMin}) can be used to design magnetic traps for
specific purposes. Here we present an IP trap that is optimized
for high longitudinal oscillation frequency. Applying equation
\ref{eq:BWX} to the H-shaped IP trap
(fig.\,\ref{fig:ContoursH}~d-e), simple calculations show that,
for a given height $z_0$, the longitudinal field curvature is
maximized if the intersection points are placed at
\begin{equation}
 x_{1,2}=\mp z_0
\end{equation}
from the center.

\begin{figure}
    \centering
  \includegraphics[width=\columnwidth,keepaspectratio]{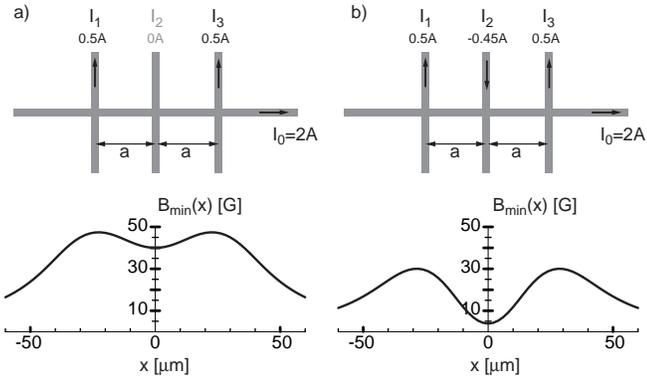}
 \caption{Longitudinal trapping field for $B_{0,y}=160$\,G. a) The H-shaped IP-trap with $a=z_0$ is optimized for maximum field
 curvature along the slow axis. b) An opposed current in the
 center of the to intersecting wires
 allows a further
 increase of the longitudinal field curvature and trapping
 frequency. The modulus of current $I_2$ must be smaller than
 $\frac{I_1+I_3}2$ in order to have a field minimum different from
 zero.
}
 \label{fig:BugTrap}
\end{figure}

Fig.\,\ref{fig:BugTrap} (a) illustrates that the center field of
this wire configuration attains comparably high values. A third
current crossing the trap axis in the center of the two parallel wires
and in opposite direction can thus be used to further increase the
field curvature and to thereby decrease the absolute value of the
minimum field strength.

The potential of this optimized
four-wire trap is shown in fig.\,\ref{fig:BugTrap}~(b).
Table~\ref{tab:trap} lists, for all three eigenaxes, the field curvature,
the oscillation frequency $\nu$ and the Lamb-Dicke parameter defined
as $(\nu\sub{r}/\nu)^2$, where $\nu\sub{r}=(\hbar
k)^2/(2m)$ is the recoil frequency for $^{87}$Rb. The field curvatures
along the eigen-axes all exceed $10^7$\,G/cm$^2$, so that $^{87}$Rb
atoms in the \ket{F=2,m=2}-ground state can be trapped within the
Lamb-Dicke regime along all axes.

\begin{table}
\caption{Parameters of the optimized four-wire trap.}
\label{tab:trap}
\begin{center}
\begin{tabular}{c|c|c|c}
&&&Lamb-Dicke\\
  dir.  & field curvature $\kappa$ & osc. freq. $\nu$ &parameter \\
\hline  1         & $1.09\cdot10^9$\,G/cm$^2$ & 42\,kHz & 0.30\\
  2         & $1.09\cdot10^9$\,G/cm$^2$ & 42\,kHz & 0.30\\
  3         & $1.63\cdot10^7$\,G/cm$^2$ & 5.1\,kHz & 0.86
\end{tabular}
\end{center}
\end{table}

\section{Key techniques of the microtrap experiment}
\subsection{Substrate technology}
\label{sec:substrate}

A wide variety of micro- and nanofabrication techniques appear
suitable for atom chips, including standard microelectronics
processes. Microchip materials generally have low outgassing rates,
making them suitable for UHV use, although some restrictions arise
from the requirement of bakeability. High field gradients and
curvatures are among the most attractive features of magnetic
microtraps; in order to optimize these features, the atom chip should
provide low-resistance conductors suitable for high current densities.
Therefore, gold and copper are desirable conductor materials. Finally,
the size scale of the structures also enters in determining the most
suitable process. Pure lithographic techniques used with sputtered or
evaporated metal layers usually limit the maximum layer thickness to a
few micrometers. If the scale of the conductor width is rather in the
$10\,\mu$m range, as is the case in all current experiments, an
additional electroplating step is helpful to increase the height of
the conductor cross section, thereby lowering the resistance. {\it
  Thin-film hybrid technology} is a standard process combining
lithography and electroplating. It employs gold conductors on aluminum
oxide, aluminum nitride or sapphire substrates. The minimum width of
conductors and gaps is typically 10\,$\mu$m. We have chosen this
process for our experiments because it fulfills all the cited
requirements and because custom-made hybrids are commercially
available. A very similar, custom-developed process was used in the
atom guiding experiments at Harvard \cite{Dekker00}. One interesting
alternative is the process used by the Innsbruck group, which uses a
uniform gold layer on a gallium arsenide substrate. This custom-made
process is more difficult to produce, but has the potential to create
smaller structures and leads to a very smooth gold surface.

\begin{figure}
  \centerline{\includegraphics{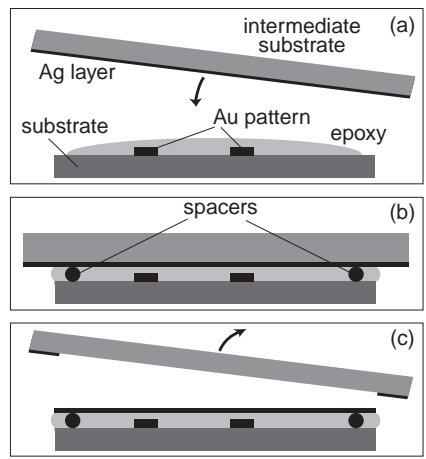}}
\caption{Steps of the replica technique leading to a smooth mirror
  surface. (a) A small amount of UHV-compatible epoxy glue (Epo-Tek
  353ND) is dispensed on the hybrid substrate; a 250\,nm silver layer (which
  will serve as the mirror) has previously been sputtered onto an
  intermediate substrate. (b) The intermediate substrate is sandwiched
  onto the hybrid substrate, with the silver layer facing the
  epoxy. Short stretches of 25\,$\mu$m-diameter gold wire serve as
  spacers. (c) After curing the epoxy, the intermediate substrate is
  lifted off, leaving the silver mirror layer on the epoxy.}
\label{fig:replica}
\end{figure}

\begin{figure}
  \centerline{\includegraphics{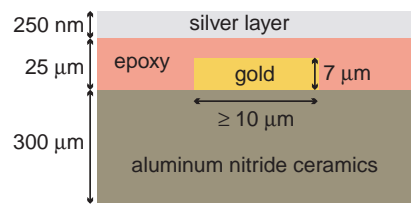}}
\caption{Layer structure of the microtrap substrate.}
\label{fig:sandwich}
\end{figure}

The mirror-MOT requires a reflective substrate surface. As the
conductor pattern forms a relief structure on the surface, directly
applying a coating to it would lead to a poor mirror with strong
diffusion from the conductor edges. We therefore use a simple
``replica optics'' procedure as described in figure~\ref{fig:replica}
to obtain a flat mirror.  The final substrate has the layer structure
shown in figure~\ref{fig:sandwich}. The size of our substrates is
$22.4\,\mbox{mm}\times 18.4\,\mbox{mm}$. This is large enough even for
complicated conductor patterns, but limits the size of the mirror and
thus the MOT beam diameter, see section~\ref{sec:loading}. Thin-film
hybrid substrates are available in larger sizes, e.g.
$2\,\mbox{in}\times 2\,\mbox{in}$.

The maximum sustainable current $I$ through the lithographic
conducters is a key parameter in order to obtain steep traps. The
thinnest wires on our hybrid substrates have a cross section of
$h=7\,\mu$m height and $w=10\,\mu$m width. With the literature value
of $\rho=2.2\,\mu\Omega$cm for the specific resistance of gold, this
leads to a resistance of 3.1\,$\Omega/$cm and a dissipated power of
$P=3.1$\,W$/$cm at $I=1\,$A. Two effects impose limitations on $I$:
global heating of the substrate by this considerable total power, and
local effects which lead to melting or evaporation of the wire at one
specific point. The first limitation can always be alleviated by
improving the heat conductivity from the substrate to the reservoir to
which it is connected, or, if necessary, by cooling this reservoir
(e.g., by using a small liquid nitrogen tank). The second, local
effect is characteristic of the microfabrication process and materials
and is the one which actually limits $I$.

In our setup, active cooling proved to be unnecessary and the cooling
system is most simple. Heat removal from the substrate is ensured by
fixing it on a copper block with an UHV-compatible, thermally
conductive epoxy resin (Epo-Tek H77). The copper block acts as a heat
buffer and is itself connected to the stainless steel vacuum system by
copper rods. We have determined the maximum sustainable currents in
hybrid wires ($h=7\,\mu$m, $w=10\,\mu$, length $l=20\,$mm) in this
assembly.  The tests were done in air with the copper block posed on a
wooden table and initially at room temperature. We obtained the
following result: currents of $I=3\,$A could repeatably be applied
for one minute; the copper block heated up by about 40\,K during that
period. At currents above $I=3.2\,$A, some wires blew up, forming
$\sim 10\,\mu$m long gaps in the wire. Thus, the highest sustainable
current density (at $I=3\,$A) was
$j=4.6\dip{6}\,\mbox{A}/\mbox{cm}^2$. The calculated magnetic field of
a $10\,\mu$m wide conductor carrying a current of $I=3\,$A is 425\,G
at a distance $d=10\mu m$ from the surface and has a gradient of
$b=4.12\cdot10^5$\,G/cm.  Superposing a sufficient longitudinal bias
field to suppress Majorana losses ($\omega\sub{osc}\approx
10\,\omega\sub{prec}$, see e.g.\ \cite{Gov00}), a Ioffe-Pritchard trap
with a transverse oscillation frequency of 270\,kHz could be
realized for $^{87}$Rb atoms. The resulting confinement leads to a
Lamb-Dicke parameter of 0.12 with respect to the D2 line,
corresponding to a ground state $1/e^2$ diameter of 60\,nm.  These
surprising figures suggest that traps with exciting new properties may
be realized with this simple technique.

\subsection{Vacuum system and outgassing issues}
The conductors on the substrate are contacted to connector pins by
gold wire bonding. This technique, extensively used in microchip
packaging, requires no soldering compounds and is therefore well
suited for UHV use. However, a special wire bonder is required.
Contacting with a silver-filled epoxy is an alternative if a wire
bonder is not available \cite{Folman00}. Kapton isolated copper wires
with matching UHV connectors are used between the connector pins and
the vacuum feedthrough.

Considering the quantity of materials employed {\it in vacuo}, the
final pressure in the system is of interest.  Our small-volume
vacuum system is pumped by a 25\,l$/$s ion pump and a simple titanium
sublimator. Both pumps are connected to the glass cell containing the
substrate by 35\,mm stainless steel tubing, involving several
$90^\circ$ angles between the cell and the pumps. A vacuum meter,
which is located about midways between the pumps and the glass cell,
indicates a final pressure of $3\dip{-10}$\,mbar typ.\ after 3 days of
baking at $140^\circ$C. More importantly, we observe $1/e$ lifetimes
in the magnetic trap of $\tau\sim 4\dots 5\,$s, depending on the Rb
pressure. It seems reasonable to expect an order of magnitude
gain if a separate MOT chamber is introduced and the pumping speed
is improved. Thus, the atom chip materials are compatible with the
vacuum level required for evaporative cooling or certain quantum
manipulation experiments.

\subsection{Detection}
\begin{figure}
  \centerline{\includegraphics{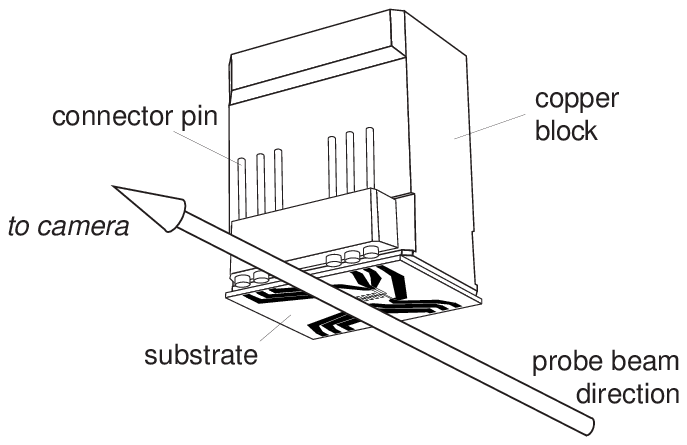}}
\caption{Geometry of the experiment. The substrated is mounted upside
  down in the vacuum chamber, so that the atoms are trapped below the
  surface. This enables time-of-flight imaging to measure the velocity
  distribution. The probe beam is directed parallel to the surface
  along the $y$ axis.}
\label{fig:probeBeam}
\end{figure}

The substrate is mounted upside down in the vacuum chamber, so that
atoms are ``hanging'' below its surface. This enables time-of-flight
imaging to measure the velocity distribution. A probe beam for
absorption imaging is directed parallel to the surface along the $y$
axis (figure \ref{fig:probeBeam}). The shadow of the atom cloud is
imaged onto a 12 bit CCD camera by a multi-element zoom lens.
Magnification is limited by the requirement to image the full $\sim
7$\,mm length of the atomic conveyer belt. This in turn sets the
resolution to 23\,$\mu$m in object space, limited by the CCD pixel
size.

After every absorption image, a reference image is taken without
atoms. Dividing the intensities of both images and taking the
logarithm yields the optical densities. All images shown in the
following section are obtained in this way.

\section{The trap loading procedure}
\label{sec:loading}

\begin{figure}
\centerline{\includegraphics{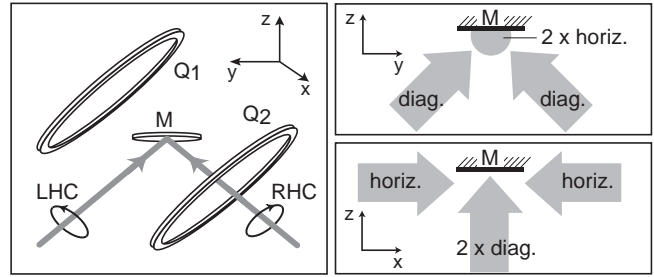}}
\caption{The mirror-MOT. Left: perspective view indicating beam
  helicities of the diagonal beams which are incident on the mirror
  $M$, and the orientation of the quadrupole coils $Q1$ and $Q2$. (The
  horizontal beams are not shown.) Right: Projections on the $yz$ and
  $xz$ planes.}
\label{fig:mirrorMOT}
\end{figure}

\begin{figure}
  \centerline{\includegraphics[width=0.65\columnwidth]{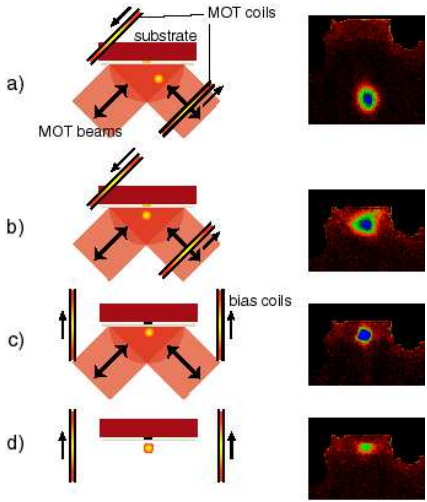}}
\caption{The loading procedure. The left column schematically depicts
  the laser beams, currents and coils employed in each step, the right
  column shows corresponding absorption images.(a) Loading of the
  mirror MOT, typ.~5\,s at a detuning $\delta=-2\,\Gamma$. (b) The MOT
  center is shifted closer to the surface by imbalancing the currents
  in the quadrupole coils (20\,ms). (c) The MOT coils are switched
  off, the MOT quadrupole field is now generated by a microtrap wire
  and bias coils. At the end of this phase, the gradient is increased
  and the repumping power reduced.  Magnetic fields are switched
  off, the atoms are further cooled in optical molasses
  ($\delta=-8\,\Gamma$, 200$\,\mu$s). A circularly polarized pumping
  pulse then spin-polarizes the atoms in a magnetic bias field
  (200\,$\mu$s). (d) The magnetic microtrap is switched on.}
\label{fig:loading}
\end{figure}

One of the key issues of an atom chip experiment is to load cold atoms
into microtraps which have much smaller volumes than usual magnetic
traps and are located in close proximity to a substrate surface. Our
loading scheme relies on a mirror-MOT (\cite{Reichel99}, see also
\cite{Pfau97}) for preparing cold atoms close to a surface, and on a
field switching procedure in the MOT phase (from a ``macroscopic''
quadrupole field to a ``microscopic'' one) to achieve reproducible
trap loading without the need for precision field alignments. A
schematic of the mirror-MOT is shown in fig.~\ref{fig:mirrorMOT}. Two
of the six MOT laser beams are generated by reflection on the mirror,
the resulting total light field is identical to that of a standard
MOT. The loading procedure has been described in \cite{Reichel99} and
is summarized in figure \ref{fig:loading}. We obtain up to $6\dip{6}$
atoms trapped in the MOT, limited by the small trapping beam $1/e^2$
diameter of $8.5$\,mm (which is itself limited by the substrate size,
as MOT beams are reflected on it), by the total laser power of $\sim
20\,$mW, and by the rubidium pressure. We adjust this pressure to
rather low values, i.e. on the same order as the
$\sim3\dip{-10}\,$mbar background pressure, in order to achieve long
magnetic trapping times.  A dispenser mounted above the substrate
serves as thermal rubidium source and allows for relatively fast
pressure adjustments.

\section{Demonstration of fundamental microtrap potentials}
\label{sec:fundamental}

\begin{figure}
  \centerline{\includegraphics{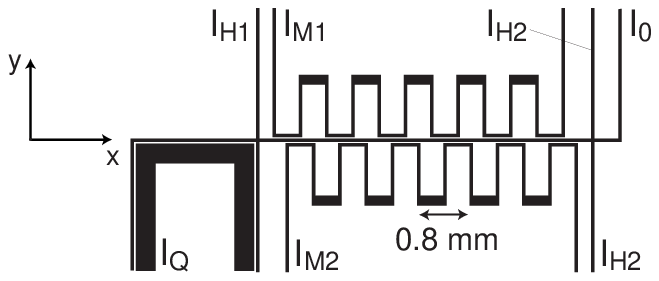}}
\caption{The complete conductor pattern used in the experiments. For
  every experiment described here, a suitable subset of this pattern
  is employed. The width of all conductors except $I_Q$ is 50\,$\mu$m.}
\label{fig:motorLayout}
\end{figure}

In Sect.~\ref{sec:design} we have described how simple conductor
patterns can be used as ``modules'', which can be combined to form
complex potentials. Some of these modules create potentials which are
useful by themselves, and we start our presentation of experimental
results with two of these: First, a conductor cross is
used to create an IP trap in which the long axis and the atomic
polarization can be rotated about the conductor intersection. In the
second experiment, a long, ``Z'' shaped conductor (which approximates
the ``H'' shape of Sect.~\ref{sec:design}) creates a very elongated IP
trap, in which the atoms are strongly confined transversally and move
almost freely about $\sim 7\,$mm along the long axis.

All experiments are carried out with the conductor pattern of
figure~\ref{fig:motorLayout}\cite{Reichel99,Haensel00a}. The variety
of basic conductor shapes which it contains makes it versatile enough
for a large number of different experiments.

\subsection{Rotatable IP trap}
\begin{figure}
  \centerline{\includegraphics[width=0.95\columnwidth]{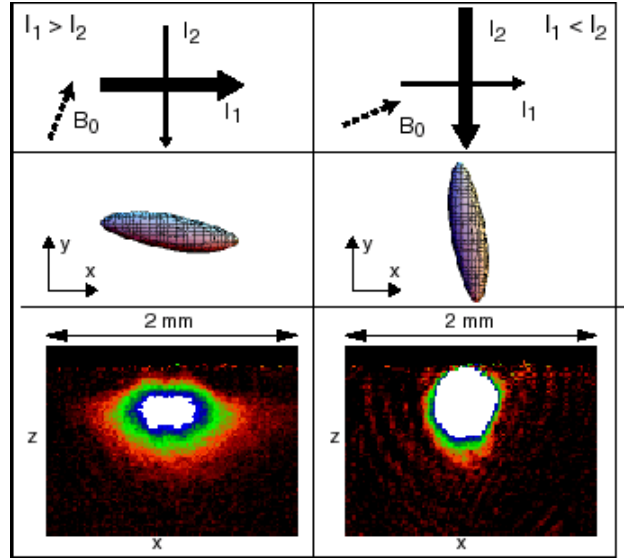}}
\caption{Ioffe-Pritchard trap created by two intersecting
  conductors. The left column corresponds to $\I{1}>\I{2}$ and
  $\abs{B_{0,y}}>\abs{B_{0,x}}$, in the right column both relations
  are reversed. Top row: Conductor pattern; the thickness of the
  arrows corresponds to the magnitude of the current. Dashed arrows
  indicate the bias field direction. Middle row: Calculated contours
  of the magnetic field modulus $\abs{B(x,y)}$ indicating how the long
  trap axis turns. The left potential continuously transforms into the
  right one when the parameters are changed smoothly. Bottom row:
  Absorption images corresponding to the two situations. The final
  trap (right) contains $4.5\dip{5}$ atoms. The trap parameters are
  given in the figure caption.}
\label{fig:crossExp}
\end{figure}

It was shown in Sect.~\ref{sec:intersection} that two crossing
conductors carrying different currents create an IP trap when a
suitably oriented bias field is superposed. As the conductor pattern
is symmetric, different orientations of the long trap axis can be
obtained by exchanging the currents and bias field components (upper
part of fig.~\ref{fig:crossExp}). By smoothly varying the currents and
bias field components, the long axis may even be turned
continuously\footnote{When a pure symmetric conductor cross is used,
  the potential passes through a quadrupole ring shape for the
  intermediate situation $I_1=I_2, \abs{B_{0,x}}=\abs{B_{0,y}}$. However,
  the additional field components introduced by the bent conductors in
  our experiment lift this degeneracy, so that the trap remains of the
  IP type during the whole rotation.} by 90$^\circ$. We use the
intersection of $\I{0}$ and $\I{H1}$ to create a trap of this type
(cf.~Fig.~\ref{fig:motorLayout}). The lower part of
Fig.~\ref{fig:crossExp} shows absorption images of the atoms in such a
trap for different orientations of the long axis. The right image
shows a trap that was obtained by turning as described above. 
The trap parameters are $\I{1}=0.2$\,A,
  $\I{2}=-1.2$\,A, $B_{0,x}=10$\,G, $B_{0,y}=4$\,G, leading to
  curvatures
  $\mathbf{\kappa}=(8.85,8.25,0.59)\dip{4}\,\mbox{G}/\mbox{cm}^2$ and
  to oscillation frequencies $\mathbf{\nu}=(378,365,97)\,$Hz.
The ability to align the
magnetic polarization of the trapped cloud can be essential when a
well-defined coupling to a polarized external field is required, such
as the evanescent light field of a whispering gallery mode in a high-finesse
microsphere resonator \cite{Treussart94,Vernooy98}.

\subsection{Highly elongated IP trap}
\begin{figure*}
  \centerline{\includegraphics{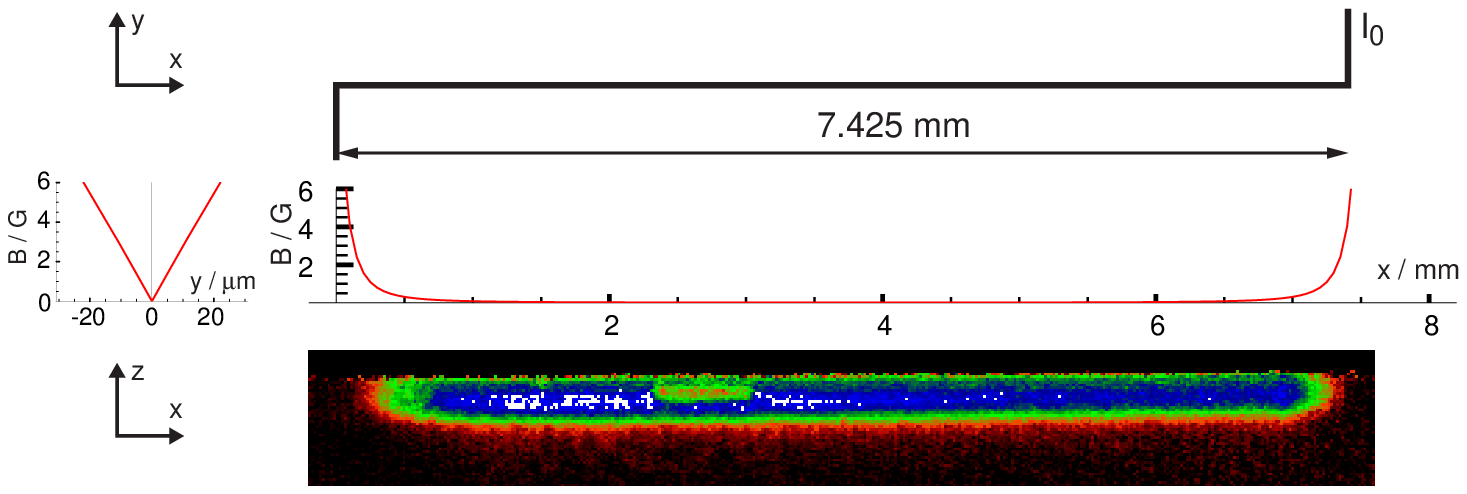}}
\caption{Ioffe-Pritchard trap with strong transverse confinement. Top:
  Conductor pattern. Center: Contours of the magnetic field modulus
  $\abs{B(x,z)}$ for the following parameters:
  $I_0=1\,A,\vect{B}=\vecy\cdot 24$\,G. Bottom: Absorption image of a cloud
  of $1.2\dip{6}$ atoms in this potential. (NB: The feature to the left
  of the cloud center is an artefact resulting from saturation of some
  atoms by a spurious reflection of the resonant probe beam.)}
\label{fig:longIoffe}
\end{figure*}

Potentials that restrict the atomic motion to one dimension are
interesting from both the theoretical and the experimental point of
view. The scattering properties of cold atoms \cite{Weiner99} change
dramatically in reduced dimensionality \cite{Olshanii98,Mandonnet00};
when the transverse level spacing exceeds the thermal energy of the
atoms, one may construct single-mode de Broglie waveguides
\cite{Thywissen99}, which are one of the building blocks of future
trapped-atom interferometers \cite{Adams94}. With lithographic
magnetic traps, the construction of such waveguide potentials is
straightforward since they derive directly from the fundamental
element of a single wire with a perpendicular bias field. As shown in
Sect.~\ref{sec:substrate}, our $10\,\mu$m thin-film conductors permit
transverse level spacings of several 100\,kHz for Rb atoms.  In the
pattern of fig.~\ref{fig:motorLayout}, $I_0$ may be used to create an
IP trap of this kind, with steep transversal confinement and an almost
box-like longitudinal potential created by the bent parts of the
conductor. The width of this conductor is $50\,\mu$m, but in our
experiment the transverse curvature is not yet limited by this wire
width. In fact, our modest bias coil systems permits a maximum bias
field of only $24\,$G, so that we cannot as yet exploit the full
potential of the conductor pattern. At $B_{0,y}=24\,$G, the potential of
fig.~\ref{fig:longIoffe} (center) is obtained. The bias field in this
example is uniquely generated by the wire sections which are parallel
to the $y$ axis, and is correspondingly small (0.01\,G in the center).
However, an external bias field along the $x$ axis is compatible with
this potential (see next section).  Figure~\ref{fig:longIoffe} shows
an absorption image in the potential described above. In this guide,
the atoms experience a transverse field gradient of 3000\,G$/$cm,
which allows for trapping frequencies up to 10\,kHz, depending on the
bias field along the trap axis. (Cf. Sect.~\ref{sec:substrate} for the
choice of the bias field).

\section{Linear cold atom collider}

Highly elongated, cigar-shaped potentials like the one of
fig.~\ref{fig:longIoffe} have been considered by theorists in the
context of one-dimensional scattering of cold atoms, for which
surprising properties have been predicted \cite{Olshanii98,Mandonnet00}. Cold
collisions are also a key ingredient of quantum computing schemes with
cold atoms \cite{Calarco00}. An ideal situation consists of two atoms
or atom clouds strongly confined in two dimensions and freely moving
along the third, approaching each other with well-defined initial
momentum. Such a situation can be realized in the device we
demonstrate in this section. It separates two samples of cold atoms so
that they move to the ends of the cigar-shaped potential, and then
releases them with opposite linear velocities (while maintaining the
transverse confinement), so that they meet in the center of the
trap. Although our atomic samples, which have not been evaporatively
cooled, are still too hot to enter into the one-dimensional regime,
this device is a good illustration of the atom manipulation
capabilities offered by lithographic microtraps.

\begin{figure}
  \centerline{\includegraphics{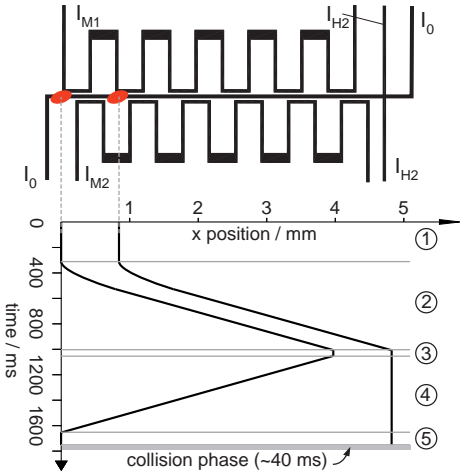}}
\caption{Top: Wire configuration used for the atomic conveyer 
  belt. The positions of the two atom clouds after transfer from the
  mirror-MOT are indicated. Bottom: collider preparation scheme: 1)
  After loading the two leftmost potentials from the MOT, a radio
  frequency is applied for 300\,ms to remove hot atoms which spill
  over into the adjacent potentials. 2) The conveyer belt is running:
  all magnetic minima are smoothly shifted to the right. 3) When the
  right cloud has arrived at its destination point, the current
  through H2 is increased. This leads to a decrease of the right
  cloud's trap potential so that this cloud remains stationary
  independently of the shifting process (see
  fig.~\ref{fig:colliderPotential}). 4) The left cloud is brought back
  to its starting point.  5) 100~ms of radio frequency remove hot
  atoms from the potentials in between both atom clouds. After this
  last preparation step, both clouds are released (see
  fig.~\ref{fig:colliderPotential}). The actual interaction phase in
  which the two clouds overlap takes less than 10\,ms.}
\label{fig:colliderPreparation}
\end{figure}

\begin{figure}
  \centerline{\includegraphics[width=0.95\columnwidth]{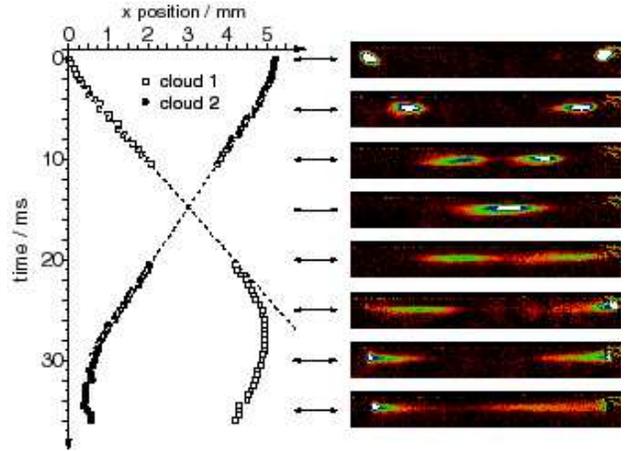}}
\caption{Linear collider experiment. Left: time evolution of the
  centers of mass of the two clouds.  Right: Absorption images.}
\label{fig:colliderResult}
\end{figure}

The device is based upon the atomic conveyer belt which we recently
demonstrated \cite{Haensel00a}, and the potential of which is which is
described in \cite{Reichel99}. It enables us to cut a long Ioffe trap
into a chain of small, tightly confined Ioffe traps, which can be
moved along the central wire axis by smoothly alternating the current
flowing through the additional wires M1 and M2 (see
fig.~\ref{fig:colliderPreparation}).  To separate two atom clouds
widely, we load the two leftmost magnetic conveyer belt potentials and
bring them into the collider starting point by the procedure described
in fig.~\ref{fig:colliderPreparation}. Next, we abruptly turn off all
wire currents except the central one. As described in
Sect.~\ref{sec:fundamental}, the atoms are now almost free along the
long axis and tightly confined in transverse direction. Due to the
steep increase of the longitudinal potential at its ends, the atoms
are accelerated towards each other. Fig.~\ref{fig:colliderResult}
shows the center of mass (CM) position of both clouds before and after
their encounter. The dashed lines are linear fits on the last ten points
before the clouds overlap. As the kinetic energy of the clouds is much
larger than the transverse level spacing, collisions will lead to a
deviation from the fit line of the CM positions after the encounter
due to the redistribution of kinetic energy into the transverse
degrees of freedom. In the present demonstration experiment, which was
not optimized for high densities, the 10\,ms encounter does not lead
to a significant effect of collisions. Nevertheless it represents
probably the most complex manipulation of atom clouds that has been
carried out in magnetic traps, and thus shows the power of the new and
flexible manipulation techniques which lithographic microtraps have to
offer.

\begin{figure}
  \centerline{\includegraphics{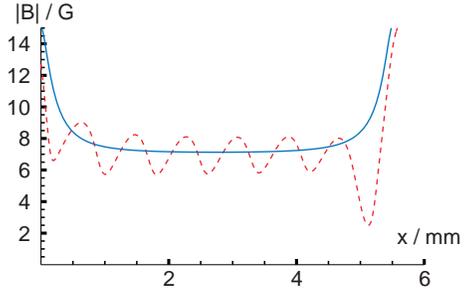}}
\caption{Potential for linear collision of two clouds. Dashed line: before
  release; solid line: after release.}
\label{fig:colliderPotential}
\end{figure}

\section{Outlook}
The results we have presented show that lithographic microtraps offer
truly intriguing possibilities for the controlled manipulation of
cold atoms. Multiple atom samples can be moved independently in close
proximity to a surface while remaining tightly confined. Complicated
potentials can easily be built from simple elements such as the wire
cross. A logical next step will be to evaporatively cool the atoms in
such a trap, a task for which microtraps are well suited due to the
high compression which they allow one to achieve. If Bose
condensation, or indeed preferential population of any single
vibrational level can be achieved, a host of novel applications
becomes accessible, ranging from trapped-atom interferometry to
quantum gate operations employing controlled collisions.

%

\begin{thebibliography}{10}

\bibitem{Ketterle99a}
{See for example W. Ketterle, D.~S. Durfee, and D.~M. Stamper-Kurn, in {\em
  Proceedings of the International Scool of Physics ``Enrico Fermi''}, edited
  by M. Inguscio, S. Stringari, and C. Wieman, cond-mat/9904034 (1999).}

\bibitem{Pinkse00}
P.~W.~H. Pinkse, T. Fischer, P. Maunz, and G. Rempe, Nature {\bf 404},  365
  (2000).

\bibitem{Ye99}
J. Ye, D.~W. Vernooy, and H.~J. Kimble, Phys.~Rev.~Lett. {\bf 83},  4987
  (1999).

\bibitem{Calarco00}
T. Calarco, E.~A. Hinds, D. Jaksch, J. Schmiedmayer, J.~I. Cirac, and P.
  Zoller, Phys.~Rev.~A {\bf 61},  022304  (2000).

\bibitem{Weinstein95}
J.~D. Weinstein and K.~G. Libbrecht, Phys.~Rev.~A {\bf 52},  4004  (1995).

\bibitem{Reichel99}
J. Reichel, W. H{\"a}nsel, and T.~W. H{\"a}nsch, Phys.~Rev.~Lett. {\bf 83},
  3398  (1999).

\bibitem{Folman00}
R. Folman, P. Kr{\"u}ger, D. Cassettari, B. Hessmo, T. Maier, and J.
  Schmiedmayer, Phys.~Rev.~Lett. {\bf 84},  4749  (2000).

\bibitem{Rosenbusch00}
P. Rosenbusch, B.~V. Hall, I.~G. Hughes, C.~V. Saba, and E.~A. Hinds,
  Phys.~Rev.~A {\bf 61},  031404(R)  (2000).

\bibitem{Mueller99}
D. M{\"u}ller, D.~Z. Anderson, R.~J. Grow, P.~D.~D. Schwindt, and E.~A.
  Cornell, Phys.~Rev.~Lett. {\bf 83},  5194  (1999).

\bibitem{Dekker00}
N.~H. Dekker, C.~S. Lee, V. Lorent, J.~H. Thywissen, S.~P. Smith, M.
  Drndi{\'c}, R.~M. Westervelt, and M. Prentiss, Phys.~Rev.~Lett. {\bf 84},
  1124  (2000).

\bibitem{Haensel00a}
W. H{\"a}nsel, J. Reichel, P. Hommelhoff, and T.~W. H{\"a}nsch,
  quant-ph/0008111  (2000).

\bibitem{Cassettari00}
D. Cassettari, B. Hessmo, R. Folman, T. Maier, and J. Schmiedmayer,
  quant-ph/0003135  (2000).

\bibitem{Mueller00}
D. M{\"u}ller, E.~A. Cornell, M. Prevedelli, P.~D.~D. Schwindt, A. Zozulya, and
  D.~Z. Anderson, physics/0003091  (2000).

\bibitem{Fortagh98}
J. Fortagh, A. Grossmann, C. Zimmermann, and T.~W. H{\"a}nsch, Phys.~Rev.~Lett.
  {\bf 81},  5310  (1998).

\bibitem{Denschlag99}
J. Denschlag, D. Cassettari, and J. Schmiedmayer, Phys.~Rev.~Lett. {\bf 82},
  2014  (1999).

\bibitem{Thywissen99}
J.~H. Thywissen, M. Olshanii, G. Zabow, M. Drndi{\'c}, K.~S. Johnson, R.~M.
  Westervelt, and M. Prentiss, Eur.~Phys.~J.~D {\bf 7},  361  (1999).

\bibitem{Gov00}
S. Gov, S. Shtrikman, and H. Thomas, J.~Appl.~Phys.~D {\bf 87},  3989  (2000).

\bibitem{Pfau97}
T. Pfau and J. Mlynek, "OSA Trends in Optics and Photonics" {\bf 7},  33
  (1997).

\bibitem{Treussart94}
F. Treussart, J. Hare, L. Collot, V. Lef{\`e}vre, D.~S. Weiss, V. Sandoghdar,
  J.~M. Raimond, and S. Haroche, Optics~Lett. {\bf 19},  1651  (1994).

\bibitem{Vernooy98}
D.~W. Vernooy, A. Furusawa, N.~P. Georgiades, V.~S. Ilchenko, and H.~J. Kimble,
  Phys.~Rev.~A {\bf 57},  R2293  (1998).

\bibitem{Weiner99}
J. Weiner, Rev.~Mod.~Phys. {\bf 71},  1  (1999).

\bibitem{Olshanii98}
M. Olshanii, Phys.~Rev.~Lett. {\bf 81},  938  (1998).

\bibitem{Mandonnet00}
E. Mandonnet, A. Minguzzi, R. Dum, I. Carusotto, Y. Castin, and J. Dalibard,
  Eur.~Phys.~J.~D {\bf 10},  9  (2000).

\bibitem{Adams94}
C.~S. Adams, M. Sigel, and J. Mlynek, Phys.~Rep. {\bf 240},  143  (1994).

\end{thebibliography}

\end{document}